\documentclass[twocolumn,english,prl]{revtex4}
\usepackage[T1]{fontenc}
\usepackage[latin9]{inputenc}
\usepackage{graphicx}
\usepackage{amsfonts}

\makeatletter
\@ifundefined{textcolor}{}
{%
 \definecolor{BLACK}{gray}{0}
 \definecolor{WHITE}{gray}{1}
 \definecolor{RED}{rgb}{1,0,0}
 \definecolor{GREEN}{rgb}{0,1,0}
 \definecolor{BLUE}{rgb}{0,0,1}
 \definecolor{CYAN}{cmyk}{1,0,0,0}
 \definecolor{MAGENTA}{cmyk}{0,1,0,0}
 \definecolor{YELLOW}{cmyk}{0,0,1,0}
}

\@ifundefined{definecolor}
 {\usepackage{color}}{}
\@ifundefined{definecolor}
 {\@ifundefined{definecolor}
 {\usepackage{color}}{}
}{}
\@ifundefined{definecolor}
 {\@ifundefined{definecolor}
 {\@ifundefined{definecolor}
 {\usepackage{color}}{}
}{}
}{}
\@ifundefined{definecolor}
 {\@ifundefined{definecolor}
 {\@ifundefined{definecolor}
 {\@ifundefined{definecolor}
 {\usepackage{color}}{}
}{}
}{}
}{}

\makeatother

\usepackage{babel}

\usepackage{amsmath}

\makeatother

\usepackage{babel}

\makeatother

\usepackage{babel}

\usepackage{braket}

\makeatother

\usepackage{babel}
\begin{document}

\title{Quantum Logic  with Interacting Bosons in 1D}

\author{Yoav Lahini$^{1}$}

\email{lahini@mit.edu}

\author{Gregory R. Steinbrecher$^{2}$}

\author{Adam D. Bookatz$^{1}$}

\author{Dirk Englund$^{2}$ }

\affiliation{$^{1}$Department of Physics, Massachusetts Institute of Technology,
Cambridge MA, USA}

\affiliation{$^{2}$Research Laboratory of Electronics, Department of Electrical
Engineering and Computer Science, Massachusetts Institute of Technology,
Cambridge MA, USA}
\begin{abstract}
We present a scheme for implementing high-fidelity quantum logic gates
using the quantum walk of a few interacting bosons on a one-dimensional
lattice. The gate operation is carried out by a single compact lattice
described by a one-dimensional Bose-Hubbard model with only nearest-neighbor
hopping and on-site interactions. We find high-fidelity deterministic
logic operations for a gate set (including the CNOT gate) that is
universal for quantum information processing. We discuss the applicability
of this scheme in light of recent developments in controlling and
monitoring cold-atoms in optical lattices, as well as an implementation
with realistic nonlinear quantum photonic devices. 

\end{abstract}
\maketitle
\emph{Introduction \textemdash{} }Quantum walks (QWs) are unitary
processes that describe the quantum-mechanical analogue of the classical
random walk process \cite{aharonov_quantum_1993,farhi_quantum_1998,kempe_quantum_2003}.
Since their conception, there has been a broad interest in their possible
use for quantum information processing \cite{kempe_quantum_2003,salvador_elias_venegas-andraca_quantum_2012}.
Two mathematical models for QWs have been developed: the discrete-time
QW \cite{aharonov_quantum_1993}, in which the particle takes discrete
steps in a direction given by a dynamic internal degree of freedom
(a coin), and the continuous-time QW \cite{farhi_quantum_1998} in
which the dynamics are described by Hamiltonian evolution on a lattice
in the tight-binding representation. Here, we consider the continuous-time
quantum walk on a one-dimensional lattice. 

Experimentally, QWs have been implemented with photons \cite{do_experimental_2005,perets_realization_2008,bromberg_quantum_2009,peruzzo_quantum_2010,broome_discrete_2010,schreiber_photons_2010,regensburger_photon_2011,rohde_multi-walker_2011},
trapped ions \cite{schmitz_quantum_2009,zahringer_realization_2010},
and ultra-cold atoms \cite{weitenberg_single-spin_2011,fukuhara_microscopic_2013,2014arXiv1409.3100P},
among other platforms. Specifically in the field of ultra-cold atoms
\cite{lahini_quantum_2012,2014arXiv1409.3100P}, the degree of experimental
control is remarkable: it is possible to prepare an initial state
with single-site and single-particle resolution, to create arbitrary
one- or two-dimensional lattice potentials, to determine the interaction
between the particles, and to directly monitor in real space the evolving
many-body distribution. 

The ability to control and monitor quantum particles with such precision
offers an interesting route to the implementation of quantum information
processing and quantum computation schemes. 
Universal quantum computation has been theoretically shown possible using QWs with 
interacting particles on certain non-trivial two-dimensional lattices \cite{childs_universal_2013,PhysRevA.85.052314}
and on one-dimensional lattices with a large number of degrees of freedom at each lattice site \cite{aharonov2009power, chase_and_landahl_2008, PhysRevA.78.032311}.
However, an implementation using quantum particles hopping on a simple one-dimensional lattice, without any additional degrees of freedom, is not known.
Such a geometry would greatly simplify
the experimental implementation, bringing it into the realm of recently
reported experimental techniques \cite{2014arXiv1409.3100P}. Furthermore,
one-dimensional implementations offer other important practical advantages,
for example freeing the second spatial dimension for important tasks
such as error correction or connecting remote qubits. Other
important tasks such as process tomography could still be performed
in 1D (see Supplementary Information). 

\begin{figure}
\includegraphics[clip,width=2.2in]{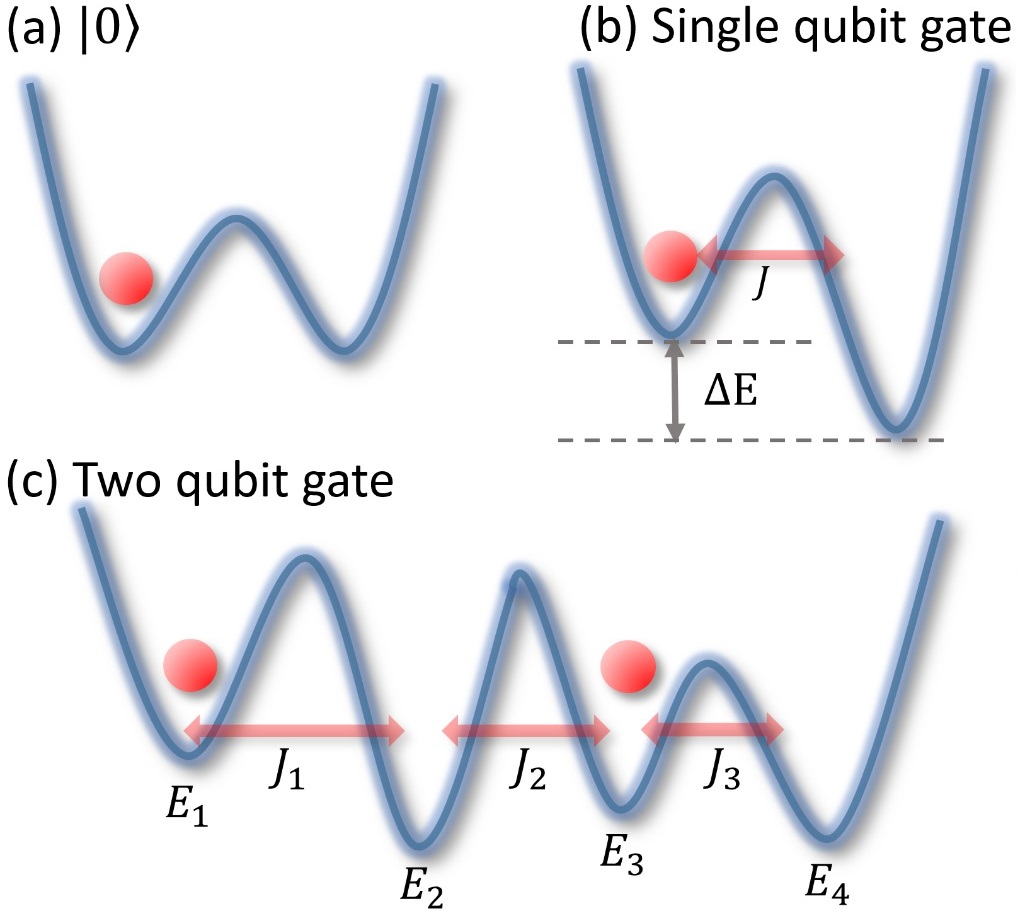}

\protect\caption{\label{fig: 1}(color online). Illustration of one-dimensional quantum
walk based quantum gates. (a) A qubit in the state $|0\rangle$ with
dual-rail encoding. (b) Implementation of a single-qubit gate. (c)
Schematic of a two-qubit system on a lattice.}
\end{figure}

In this work, we show how it is possible to use multi-particle continuous-time
quantum walks in a simple geometry \emph{\textemdash{}} a one-dimensional
lattice with only nearest-neighbor hopping and on-site interactions
\emph{\textemdash{}} as a compact platform for implementing quantum
logic. We demonstrate our approach by detailing a set of lattice potentials
that yield, with high fidelity, a universal set of quantum gates with
only two sites per qubit. Moreover, the required lattice potential
for each gate is time-invariant, simplifying the experimental implementation
and possibly reducing the total operation time. Thus, high-fidelity
gates can be constructed with lower probabilities of qubit-loss errors
(i.e. lost particles during the computation) that necessitate computationally
costly error-correction procedures. While we focus on interacting
ultra-cold bosonic atoms trapped in an optical lattice --- a physical
system in which our results can be implemented with existing experimental
techniques \cite{2014arXiv1409.3100P} --- we also discuss the possibility
of extending our analysis to nonlinear quantum photonic systems.

The dynamics of bosonic particles on a lattice is described by the
time-independent many-body Bose-Hubbard Hamiltonian

\begin{equation}
H=\sum_{m}E_{m}a_{m}^{\dagger}a_{m}+\sum_{\langle l,m\rangle}J_{l,m}a_{l}^{\dagger}a_{m}+\frac{\Gamma}{2}\sum_{m}n_{m}(n_{m}-1)\ ,\label{eq:QH-1}
\end{equation}

where $E_{m}$ is the on-site energy of site $m$, $a_{m}^{\dagger}\backslash a_{m}$
is the creation\textbackslash{}annihilation operator for a boson in
site $m$, $n_{m}=a_{m}^{\dagger}a_{m}$ is the number operator, $J_{l,m}\leq0$
is the tunneling rate between nearest neighbors, and $\Gamma$ is
the on-site interaction energy that arises when two or more bosons
occupy the same site. The unitary transformation describing the evolution
of multiple quantum particles propagating on the lattice is given
by $e^{-iHt}$, where $t$ is the propagation time. The quantum logic gates discussed here will be implemented by evolving under this Hamiltonian (with a suitable choice of parameters) for some predefined time $t_{final}$  which we take to be   $t_{final}=1$. 

\emph{Defining qubits on a lattice \textemdash{}} The basic element
of interest for quantum gates of the type discussed here is the quantum
bit, or \emph{qubit}. The continuous-time quantum walk, however, is
described by the evolution of quantum particles on a lattice according
to the Hamiltonian described in Eq. \ref{eq:QH-1}. To define our
qubits on the lattice, we use a spatial encoding where a qubit is
physically implemented by a single boson in a pair of neighboring
potential wells (see Fig. 1), with the states $|0\rangle$ and $|1\rangle$
of the qubit defined by the particle being in the left or right well
(i.e. dual-rail encoding). A single quantum particle can occupy the
two sites in a superposition, encoding a qubit without the need for
additional degrees of freedom. In this way, a system of $n$ qubits
can be realized in one dimension with $n$ bosons and $2n$ lattice
sites, with one boson in the first two sites (representing the first
qubit), one boson in the next two sites (representing the second qubit),
and so forth. Note that in this geometry, many physically permitted
lattice states (e.g. those with more than one particle on the same
site) are not members of the logical space (i.e. the multi-qubit tensor-product
space). Nevertheless, we show that it is possible to engineer the
lattice parameters such that, at time  $t=1$, $U=e^{-iH}$ maps logical states only
to other logical states with high fidelity, even though states outside
this subbasis are allowed at intermediate times.

\begin{figure}
\includegraphics[clip,width=3.4in]{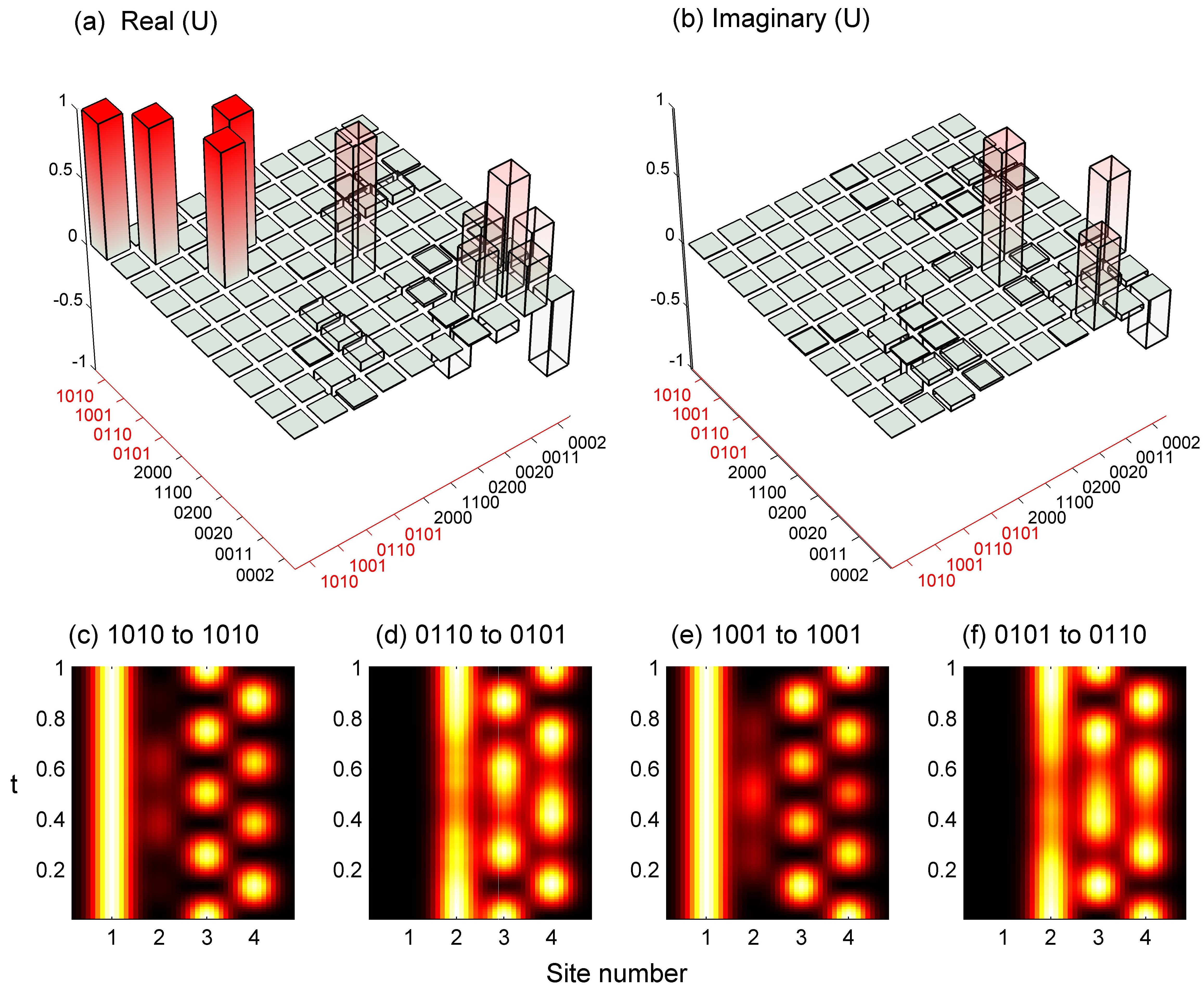}

\protect\caption{\label{fig: 2}(color online). An implementation of the controlled-NOT
(CNOT) gate according to the recipe in Eq. \ref{eq:4}. (a) The real
part and (b) the imaginary part of the two-particle unitary transform,
$U$. The CNOT gate operation corresponds to the sub-matrix of the
logic states, shown in solid-color bars and marked with red axis labels.
Plots (c)-(f) show the position (in terms of the lattice sites, 1-4)
of the particle density as a function of time, $t$, revealing
the operation principle of the gate on each logical state ($|00\rangle$,
$|10\rangle$, $|01\rangle$, and $|11\rangle$ respectively). One
observes that the target qubit (in sites 3 \& 4) performs Rabi-oscillations
that are perturbed by the state of the control qubit (in sites 1 \&
2) \emph{\textemdash{}} the target qubit performs one fewer Rabi-flip
if the control qubit is in the $|1\rangle$ state.}
\end{figure}

\emph{Implementing quantum gates \textemdash{}} Having defined our
qubits, we turn to the task of designing a universal set of quantum
gates, i.e. finding lattice parameters that yield desired unitary
transformations on the logical space. Designing and building quantum
logic gates remains one of the most difficult aspects of quantum computing,
and our case is no exception. From the physical description of a given
device \emph{\textemdash{}} in our case, the lattice parameters \emph{\textemdash{}}
it is straightforward to write down the many-particle Hamiltonian
and to calculate the unitary evolution operator $U=e^{-iH}$ that
fully describes the operation of the device. The inverse problem,
however, is hard: given a desired unitary $U$, it is difficult to
find a corresponding Hamiltonian that meets the physical and geometrical
constraints of the device, e.g. the one-dimensionality of the lattice.
Furthermore, if the logical quantum states are only a subset of the
full Hilbert space, then the quantum gate operation is only a sub-matrix
of the overall evolution operator $U$. In this case, $U$ is not
even uniquely defined by the desired gate operation. As described
below, we tackle these difficulties with a computational approach
that finds appropriate lattice parameters to approximate a given gate
operation with high fidelity. 

There are many options for the choice of a universal set of gates.
One useful choice is the gate set of the controlled-NOT
(CNOT) operation, along with either all single-qubit rotations (exactly
universal) or the phase-shift gates and the Hadamard gate (approximately
universal) \cite{nielsen_quantum_2000}. In the following, we elaborate
on the construction of the CNOT gate. Single-qubit gates involve only a single particle (i.e. no interaction
terms) and are straightforward to calculate, as we detail in the Supplementary
Information.

To design the CNOT gate (a two-qubit gate) with the dual-rail encoding,
we consider a lattice with four sites and two bosons. This problem
then is defined by eight lattice parameters: four on-site potential
terms ($E_{m}$ in Eq. \ref{eq:QH-1}), three tunneling terms ($J_{l,m}$),
and the interaction parameter ($\Gamma$). The complete two-body Hamiltonian
$H$ is described by a $10\times10$ matrix (the size of the Hilbert
space for two bosons in four modes). To perform the logical gate operation,
the system is evolved according to $U=e^{-iH}$. The CNOT gate operation
is then given by a $4\times4$ sub-matrix of $U$ over the logical
states $\ket{1010},\;\ket{1001},\;\ket{0110},\;\ket{0101}$ (presented
here in the occupation number basis); the other six basis states,
while physically allowed, are not members of the logical
basis. 

As explained above, finding the physical lattice parameters from the
desired gate is a non-trivial inverse problem. Using non-linear optimization
techniques \cite{johnson_nlopt_????,kan_stochastic_1987a,kan_stochastic_1987b,powell_bobyqa_2009}
(see Supplementary Information), we optimized the
eight parameters of the system to maximize the fidelity of the gate when acting on the logical input states
 under the constraints that the parameters represent a
physical one-dimensional lattice, i.e. that the on-site parameters
are real, that the tunneling parameters are real and non-positive
and connect only nearest-neighboring sites, and that the values of
the on-site, tunneling, and interaction terms are within experimentally
relevant bounds. Specifically, we demanded that $0\geq J_{l,m} \geq -J_{max}$, $-J_{max} \geq E_m \geq J_{max}$,
and $\Gamma \leq \Gamma_{max}$, where $J_{max}$ and $\Gamma_{max}$ are the largest
allowed tunneling rate and interaction level in the optimization protocol. In our optimization we set $J_{max}=4\pi$,
limiting the maximal number of tunneling events (or Rabi-oscillations) to
4. In practice, this experimental bound is dictated by the loss and
decoherence rate of the system, determining the maximal relevant
propagation time. We also set $\Gamma_{max}=10J_{max}$.

An example of a resulting lattice that yields the two-qubit CNOT gate
is given (to two decimal places) by 

\begin{equation}
G_{CNOT}=\pi\left(\begin{array}{cccc}
0.40 & 0 & 0 & 0\\
0 & 1.82 & -1.03 & 0\\
0 & -1.03 & -0.37 & -3.80\\
0 & 0 & -3.80 & -0.66
\end{array}\right)\label{eq:4}
\end{equation}

with interaction strength $\Gamma=21.68\pi$. Here, the diagonal
and off-diagonal entries of $G_{CNOT}$ represent the parameters $E_{m}$
and $J_{l,m}$, respectively, of the Hamiltonian $H$. Eq. \ref{eq:4}
represents a recipe for a four-site lattice that yields a CNOT gate
with fidelity of 99.6\%, whose operation is summarized in Fig. \ref{fig: 2}.

\begin{figure}
\includegraphics[clip,width=3.4in]{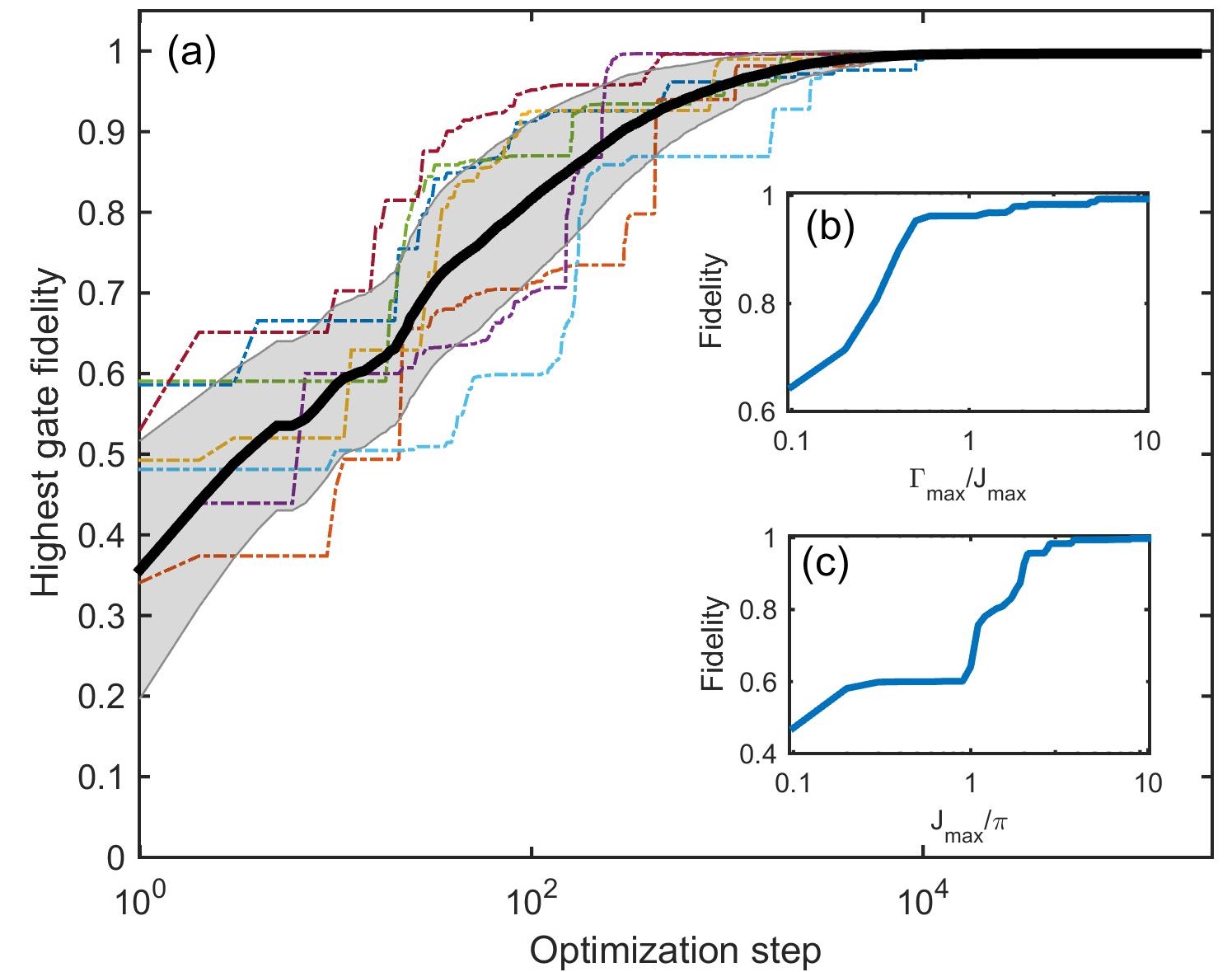}

\protect\caption{\label{fig:3}(color online). Optimization of the quantum-walk-based
CNOT gate fidelity. (a) Convergence of different optimization runs to
the optimal gate fidelity. The solid black line and the shaded area
represents the average and the standard deviation values over 512
runs. Seven example runs are shown in the background (dotted lines).
(b) Gate fidelity versus the maximum allowed interaction level $\Gamma_{max},$
at a constant $J_{max}=4\pi$. (c) Gate fidelity for different maximal tunneling
rates $J_{max}$ at a constant maximal interaction level of $\Gamma_{max} = 20\pi$.}
\end{figure}

If the bounds on the parameters are relaxed, the fidelity approaches
even closer to unity. Fig. \ref{fig:3} summarizes the optimization
results. Fig. \ref{fig:3}(a) shows the convergence of independent
runs with random starting points to the same final result. Fig. \ref{fig:3}(b)
presents the expected gate fidelity vs. the maximally allowed values
of the interaction $\Gamma_{max}$. For a fixed maximal tunneling of $J_{max}=4\pi$,
the fidelity achieves a value close to 0.95 at $\Gamma_{max}/J_{max}=0.5$
and then slowly approaches unity as this value is further increased.
In a system with a given $\Gamma_{max}$, it is still possible
to improve the fidelity further by allowing more tunneling events
to take place, i.e. increasing $J_{max}$; see Fig \ref{fig:3}(c). 

\begin{figure}
\includegraphics[clip,width=3.4in]{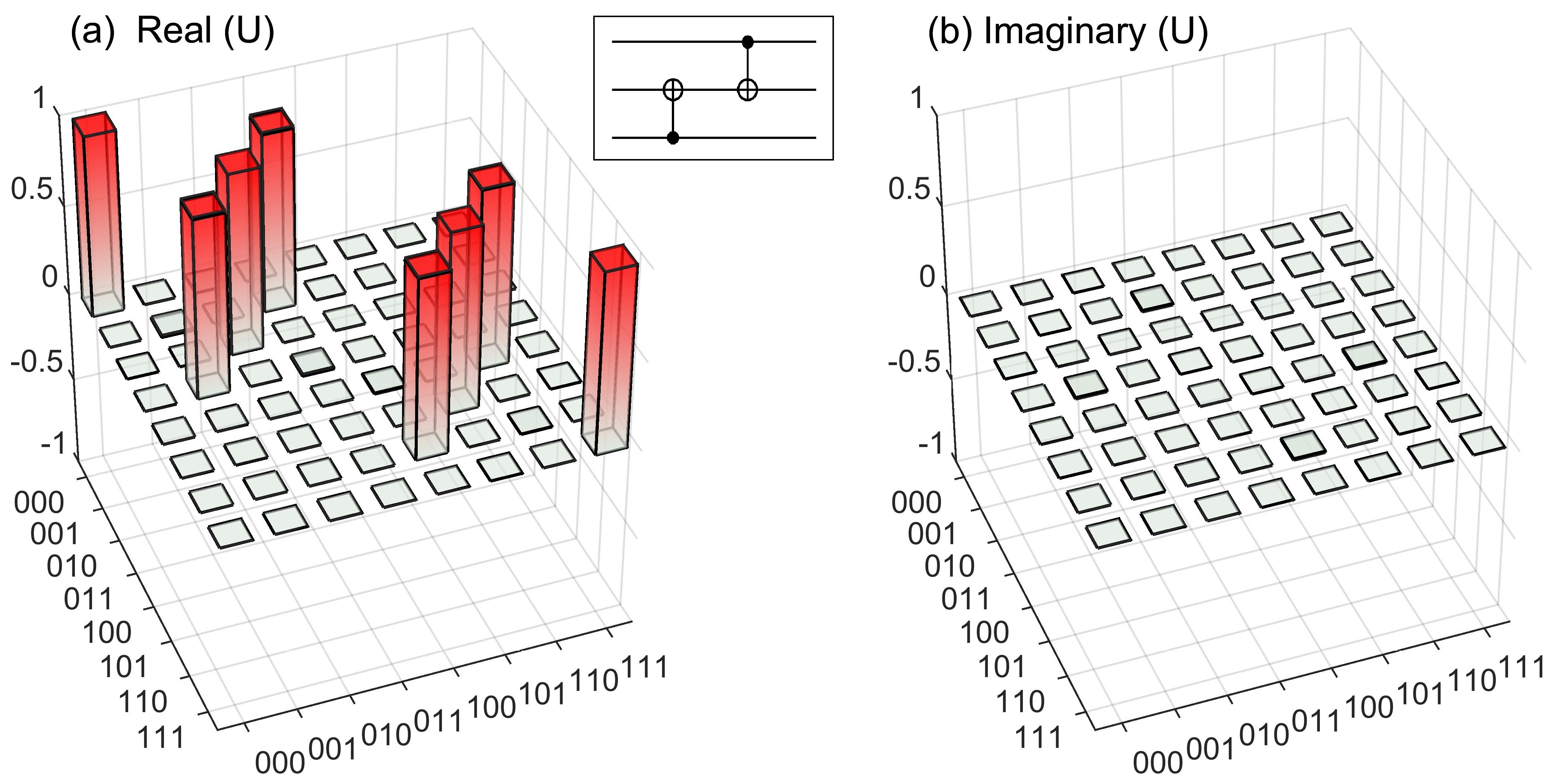}

\protect\caption{\label{fig: 4}(color online). A 3-qubit operation of 2 CNOT gates
(inset), compiled into a single gate $U$. (a) The real and (b) imaginary
parts of our implementation of $U$. Only the logical basis states
are shown.}
\end{figure}

\emph{Compiling a three-qubit primitive \textemdash{} }Implementing
a quantum algorithm using the scheme presented in this paper will
involve several lattice configurations operating in sequence, as gates
are sequentially applied in the algorithm. In principle, because the
gate set presented in this work is universal, any multi-qubit operation
can be broken down into a sequence of single- and two-qubit gates,
and thus implemented using the gates already presented. However, compiling
common multi-step operations into a single primitive based on a single,
time-independent Hamiltonian could reduce the possibility of errors
arising from dynamic changes to the lattice. As an example, we constructed
a 3-qubit gate, shown in Fig. \ref{fig: 4}. This gate is useful,
for instance, in the 2-bit Deutsch-Jozsa algorithm \cite{deutsch1992rapid},
performing the oracle for the function $f(x,y)=x\oplus y$. (All other
oracles for the 2-bit Deutsch-Jozsa algorithm are either a simple
variation of this oracle or require only single-qubit gates plus at
most one CNOT gate.) Our computational approach allowed us to find
a set of lattice parameters that realizes the complete three-qubit
operation in a single gate. Fig. \ref{fig: 4} presents an implementation
of this three-qubit operation, at a fidelity of 99.8\%, using a single,
one-dimensional six-site lattice:

\begin{equation}
G=\pi\left(\begin{array}{cccccc}
5.98 & 0 & 0 & 0 & 0 & 0\\
0 & 7.13 & -1.21 & 0 & 0 & 0\\
0 & -1.21 & 0.14 & -12.04 & 0 & 0\\
0 & 0 & -12.04 & 0.18 & -1.37 & 0\\
0 & 0 & 0 & -1.37 & 11.69 & 0\\
0 & 0 & 0 & 0 & 0 & -8.03
\end{array}\right)\label{eq:4-1}
\end{equation}

with interaction strength $\Gamma=108.24\pi$. In this case too,
the fidelity could be improved by allowing larger tunneling rates.

\emph{Photonic systems \textemdash{} }While we have focused our attention
on cold atoms, it is possible to extend our scheme to nonlinear photonic
systems, for example superconducting circuit QED systems \cite{PhysRevLett.110.060503}
or nonlinear quantum-optical systems \cite{firstenberg_attractive_2013,gullans_single-photon_2013,dayan_photon_2008,englund_controlling_2007,englund2012ultrafast,PhysRevLett.108.227402,volz2012ultrafast}.
First, consider a system described by a similar Hamiltonian
to that of Eq. 1: a set of coupled resonators exhibiting a single-photon Kerr nonlinearity \cite{kirchmair2013observation}
\begin{multline}
H=\sum_{m}\omega_{m}\left(a_{m}^{\dagger}a_{m}+\frac{1}{2}\right)\\
+\sum_{\langle l,m\rangle}J_{l,m}a_{l}^{\dagger}a_{m}+\frac{K}{2}\sum_{m}n_{m}(n_{m}-1),
\end{multline}

where $\omega_{m}$ is the resonant frequency of the $m$th cavity
and $K$ the strength of the Kerr nonlinearity. Replacing $\omega_{m}$
with $E_{m}$ and $K$ with $\Gamma$, we regain Eq. \ref{eq:QH-1}
exactly (neglecting the zero-point offset). Such systems have been
experimentally demonstrated \cite{kirchmair2013observation} and site-by-site
tunable coupled-cavity systems have been proposed \cite{gangat2013deterministic}.
However, our scheme requires a different $E_{m}$ (here, cavity frequency)
at each lattice site, which is difficult to reconcile with a high-Q
coupled resonant system. 

We can remedy this issue by instead coupling each of the cavities
to a virtual two-level atom, whose transition frequency $\omega_{a,m}$
is detuned from the (now constant) cavity resonance $\omega_{c}$
by a frequency $\Delta_{m}=\omega_{a,m}-\omega_{c}$. The Hamiltonian
is then well described by \cite{blais2004cavity} 

\begin{multline}
H=\sum_{m}\left[\omega_{c}\left(a_{m}^{\dagger}a_{m}+\frac{1}{2}\right)+\frac{\omega_{a,m}\sigma_{m}^{z}}{2}\right]\\
+\sum_{\langle l,m\rangle}J_{l,m}a_{l}^{\dagger}a_{m}+\sum_{m}g_{m}(a_{m}^{\dagger}\sigma_{m}^{-}+\sigma_{m}^{+}a_{m}),
\end{multline}

where $g_{m}$ is the atom-photon coupling rate (i.e. vacuum Rabi
frequency), and the $\sigma_{m}^{i}$ are the standard Pauli operators.
To second order in $g_{m}/\Delta_{m}$, the state of the atom induces
an a.c. Stark shift of the cavity by $\chi_{m}=g_{m}^{2}/\Delta_{m}$
\cite{schuster2007resolving}. Equivalently, a flip of the atom's
state shifts the resonance of the cavity by $2\chi_{m}$, meaning
that any photons remaining in the cavity will have acquired an extra
energy of $2\chi_{m}$. Averaged over many Rabi-oscillations, this
shifts the cavity photon frequency by $\chi_{m}$ when two photons
are on the same site, causing a faster phase evolution rate than for
a single photon. This is equivalent to the effect of an energy cost
for two photons on the same site. Thus, by tuning ${g_{m}}$ and ${\omega_{m}}$,
we have enough degrees of freedom to obtain analogous dynamics to
those of Eq. \ref{fig: 1}, having both distinct on-site energies
and a tunable nonlinear interaction.

Of particular note, the inclusion of a two-level atom provides a powerful
advantage over both cold atom systems and the aforementioned nonlinear-Kerr
cavities. As has been experimentally demonstrated \cite{schuster2007resolving},
due to the commutation of the a.c. Stark shift with the photonic and
atomic states, it is possible to perform a quantum non-demolition
measurement on the state of the atom (photons) on a given site without
affecting the photons (atom) on that site or any other states in the
system. Potentially, this could be used to monitor or tune the operation
of gates in situ or as part of an error correction scheme. 

Finally, we note that while we have focused on the capabilities of
superconducting systems in this section, devices based on quantum
dots or graphene may offer similar opportunities at optical frequencies
\cite{firstenberg_attractive_2013,gullans_single-photon_2013,dayan_photon_2008,englund_controlling_2007,englund2012ultrafast,PhysRevLett.108.227402,volz2012ultrafast}.

\emph{Conclusions \textemdash{} }We have shown how quantum logic gates
can be realized with high fidelity using the quantum walk of ultra-cold
atoms on a one-dimensional lattice under experimentally achievable
constraints. In particular, we gave a design for a high fidelity CNOT
gate along with exact descriptions of single-qubit rotations, a computationally
complete set. Additionally, we demonstrated the compilation of a higher-order
gate operation into a single operation. Our approach carries several
important advantages over previous schemes. First, due to the dual-rail
encoding we employ, the states of the system can be prepared and measured
by simply placing and detecting single atoms at certain positions,
both of which are straightforward in present experimental systems
\cite{2014arXiv1409.3100P}. Second, each quantum operation is carried
out by a single, one-dimensional, time-invariant lattice potential.
Third, the devices we propose are compact lattices of size $2n$ (where
$n$ is the number of qubits) that can be realized on a line of potential
wells with only nearest-neighbor hopping, in agreement with experimental
capabilities. Our analysis shows that similar effects could be achieved
in certain nonlinear quantum-optical systems.

\emph{Acknowledgments \textemdash{} }We acknowledge helpful discussions
with Terry Orlando, William Oliver and Markus Greiner's group. G.R.S.
was supported by the Department of Defense (DoD) through the National
Defense Science \& Engineering Graduate Fellowship (NDSEG) Program.
D.E. acknowledges support from the Sloan Research Fellowship in Physics.
Y.L. acknowledges support from the Pappalardo Fellowship in Physics.

\bibliographystyle{apsrev}
\bibliography{BH}

%%%% ---- supplement ----
%%%%%%%%%% Merge with supplemental materials %%%%%%%%%%
\pagebreak
\widetext
\begin{center}
\textbf{\large Supplementary Information: Quantum Logic with Interacting Bosons in 1D}
\end{center}
%%%%%%%%%% Merge with supplemental materials %%%%%%%%%%
%%%%%%%%%% Prefix a "S" to all equations, figures, tables and reset the counter %%%%%%%%%%
\setcounter{equation}{0}
\setcounter{figure}{0}
\setcounter{table}{0}
\setcounter{page}{1}
\makeatletter
\renewcommand{\theequation}{S\arabic{equation}}
\renewcommand{\thefigure}{S\arabic{figure}}
\renewcommand{\bibnumfmt}[1]{[S#1]}
\renewcommand{\citenumfont}[1]{S#1}

%%%%%%%%%% Prefix a "S" to all equations, figures, tables and reset the counter %%%%%%%%%%

\newcommand{\Tr}{Tr}

\newcommand{\twobytwomatrix}[4]{\left(\begin{array}{cc}
#1 & #2\\
#3 & #4
\end{array}\right)}
\renewcommand{\ket}[1]{|#1\rangle}
\newcommand{\hs}[2]{\ensuremath{\left< #1, #2 \right>_{\text{C}}}}

\emph{Single-qubit gates -}
To complement the presentation of a controlled-not (CNOT) gate in the main paper, we present the exact construction for a set of single-qubit gates that, together with the CNOT gate, make a universal quantum gate set. Since these are one-qubit operations, they are implemented using one particle in two lattice sites. As such, the interaction ($\Gamma$) is irrelevant and the matrix of lattice parameters 
\[
	G = \twobytwomatrix{E_1}{J_{12}}{J_{12}}{E_2}
\]
(with $J_{12}\leq 0$) can be directly interpreted as the Hamiltonian governing the single particle. The unitary gate, obtained by evolving with $G$ for a time $T$, is then $U=e^{-iGT}$.

\begin{figure}
\includegraphics[clip,width=3.4in]{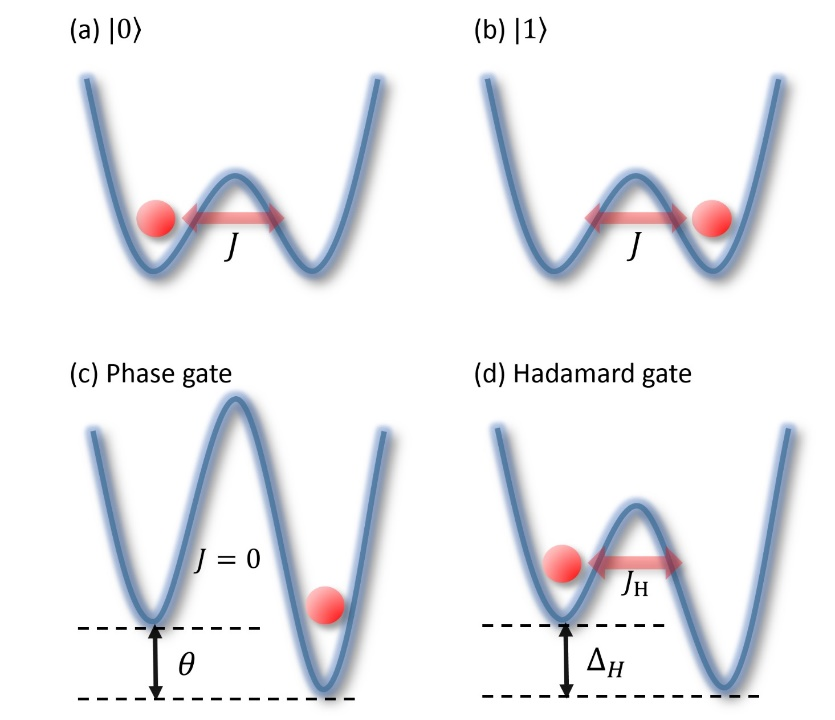}
\protect\caption{\label{fig:qubit_phase_hadamard}(color online). 
A double potential well containing a single-boson configuration representing a single qubit in the (a) $\ket{0}$ state and the  (b) $\ket{1}$ state.  (c) An implementation of the single-qubit phase gate, according to the recipe in Eq.~\ref{eq:h_phase}. (d) An implementation of the Hadamard gate, according to the recipe in Eq.~\ref{eq:h_hadamard}. 
}
\end{figure}

One simple universal quantum gate set includes the Hadamard gate, the phase-shift gate, and the controlled-NOT (CNOT) gate \cite{snielsen_quantum_2000}.
The phase-shift gate is the simplest to implement. It is composed of two decoupled lattice sites in which the on-site energy between the sites is detuned. Specifically, to implement the single-qubit phase-shift operator 
$R_\theta=\twobytwomatrix{1}{0}{0}{e^{i\theta}}$,
one may apply the single-qubit Hamiltonian
\begin{equation}\label{eq:h_phase}
	G_{R_\theta}=\twobytwomatrix{0}{0}{0}{-\theta}
\end{equation}
for time $T=1$. (Obviously a different value for $T$ may be chosen, so long as $T G_{R_\theta}$ remains the same.)
Next in complexity is the single-qubit Hadamard gate 
$H=\frac{1}{\sqrt2}\twobytwomatrix{1}{1}{1}{-1}$.
The Hamiltonian and propagation time that generates the Hadamard transformation can be solved analytically, and is given by
\begin{equation}\label{eq:h_hadamard}
	G_{H}
	=\twobytwomatrix{\sqrt2 - 1}{-1}{-1}{\sqrt2 + 1} \,,\quad T=\frac{\pi}{2\sqrt2} \,.
\end{equation}
Note that a simple tunnelling between two identical wells for half the tunnelling time (analogous to the beamsplitter operation in linear optics) will not reproduce the Hadamard gate in this case. Unlike the bulk quantum optics situation, in which the optical beamsplitter can be asymmetric in phase and thus can reproduce the Hadamard gate exactly, under our Hamiltonian dynamics the splitting is symmetric in phase and therefore modified tunnelling rates and additional diagonal terms are required in order to correct the output phases. 
We note this holds true also for integrated quantum photonics gates \cite{spoliti_silica--silicon_2008}. A summary of the physical construction for the single-qubit operations is given in Fig.~\ref{fig:qubit_phase_hadamard}.

Alternatively, one can use the (exactly) universal gate set of CNOT together with all single-qubit unitaries. Any single-qubit unitary, $U$, can be implemented by first decomposing it in the form \cite{snielsen_quantum_2000}
\[
	U = e^{i\alpha} R_z(\beta) R_x(\gamma) R_z(\delta) = e^{i\alpha} R_z(\beta) H R_z(\gamma) H R_z(\delta)
\]
where the $z$-rotation $R_z(\theta) = \twobytwomatrix{e^{-i\theta/2}}{0}{0}{e^{i\theta/2}}$ can be implemented with the Hamiltonian
\[
	G_{R_z(\theta)} = \twobytwomatrix{1}{0}{0}{-1} ,\quad T=\frac{\theta}{2} \,,
\]
and the $x$-rotation $R_x(\theta) = \exp\twobytwomatrix{0}{-i\theta/2}{-i\theta/2}{0}$ can be implemented either with the Hamiltonian
\begin{equation}\label{eq:h_Rx}
	G_{R_x(\theta)} = \twobytwomatrix{0}{-1}{-1}{0} ,\quad T=\frac{4\pi-\theta}{2}
\end{equation}
or by conjugating $R_z(\theta)$ by the Hadamard operation $H$ described earlier. The phase $e^{i\alpha}$ can be implemented with $G_\alpha = -\alpha\twobytwomatrix{1}{0}{0}{1},\, T=1$.

\newcommand{\id}{I} %identity symbol
\newcommand{\vecket}[2]{\left(\begin{array}{c}#1 \\#2\end{array}\right)}
\newcommand{\invsqrt}[1]{\frac{1}{\sqrt #1}}

\emph{Quantum process tomography for CNOT -}
The basic idea of quantum process tomography \cite{schuang_prescription_1997} is to completely determine the mathematical structure of a given operation (in our case, the CNOT on two qubits), by applying the operation to a variety of different input states and measuring each of the resulting states in a variety of bases. 

Tomography procedures allow much freedom is choosing input states and measurement bases. We want to make such choices wisely to facilitate performing the procedure on our particular experimental apparatus. Two main points guide these choices. One is that the only two-qubit gate be the purported CNOT gate itself --- all other gates required for the tomography should be relatively simple so that we adequately assess the CNOT as the major source of errors. The second is that the other gates be easy for the Bose-Hubbard model to implement, preferably with as few gates as possible (since limitations in coherency times may limit the number of gates reasonably performed using current technology). Here we suggest input states and measurement bases to accomplish these goals,
requiring only two single-qubit Bose-Hubbard Hamiltonians
 (one prior to the CNOT, and one following it) per (computational) measurement.  Specifically, each measurement in the tomography involves applying an operation of the form 
$(V_{3}\otimes V_{4})\ \text{CNOT}\ (V_{1}\otimes V_{2})$
 to a computational basis state, followed by a measurement in the computational basis, where each $V_i$ can be performed with a single one-qubit Bose-Hubbard Hamiltonian; we therefore need only keep the state coherent for three consecutive operations.

We assume that the reader is following the quantum process tomography procedure outlined in Ref.~\cite{snielsen_quantum_2000}, which requires measuring Pauli observables.
To measure a Pauli matrix $\sigma \in \lbrace \id, X, Y, Z\rbrace$, one must perform a basis-transformation gate to their output state so as to measure in the eigenbasis of $\sigma$. For $\sigma=\id$ and $Z$, this is trivial, as the computational basis suffices; however, 
for $\sigma=X$ and $Y$, we recommend measuring in the following eigenbases, which can be performed by first applying the following basis-transformation matrices,
\begin{equation}\label{eq:Ux}
	X: \left\lbrace \invsqrt2\vecket{1}{1}, \invsqrt2\vecket{1}{-1} \right\rbrace, \quad U_X=\invsqrt2\twobytwomatrix{1}{1}{1}{-1}
\end{equation}
and
\begin{equation}\label{eq:Uy}
	Y: \left\lbrace \invsqrt2\vecket{1}{i}, \invsqrt2\vecket{i}{1} \right\rbrace, \quad U_Y=\invsqrt2\twobytwomatrix{1}{-i}{-i}{1} \,.
\end{equation}
Observe that $U_X$ is the Hadamard matrix, the Hamiltonian for which was given in Eq.~\ref{eq:h_hadamard}, and that $U_Y=R_x(\pi/2)$, which is generated by the Hamiltonian in Eq.~\ref{eq:h_Rx} with $\theta=\pi/2$. Note that for $Y$ we chose a non-standard orthonormal eigenbasis so that the basis-transformation can be accomplished using a single Bose-Hubbard gate .

The input states required to perform the quantum process tomography procedure of Ref.~\cite{snielsen_quantum_2000} are straightforward to produce. For example, one may use tensor products of the following single-qubit input states:
the computational basis states $\ket{0}$ and $\ket{1}$, the $+1$ $X$-eigenvector $\ket{+} = U_X \ket{0}=\frac{1}{\sqrt2}\vecket{1}{1}$, and the $+1$ $Y$-eigenvector $\ket{-} = U_Y^\dag \ket{0}=\frac{1}{\sqrt2}\vecket{1}{i}$, where $U_X$ and $U_Y$ are given in Eqs.~(\ref{eq:Ux}) and (\ref{eq:Uy}). Note that $U_Y^\dag=R_x(7\pi/2)$ can be generated by the Hamiltonian in Eq.~\ref{eq:h_Rx} with $\theta=7\pi/2$. 

\emph{Computational Methods -}
In this section, we detail the numerical methods used to find the gates presented in the main paper. We made extensive use of a free-software implementation of a variety of numerical optimization algorithms \cite{sjohnson_nlopt_????}. This enabled us to, with a single specification of cost function and constraints, compare the success and computational cost of a number of different optimization approaches. We discovered that a randomly-seeded global optimization algorithm \cite{skan_stochastic_1987a,skan_stochastic_1987b} combined with a gradient-free local algorithm \cite{spowell_bobyqa_2009} gave the best performance, both in terms of number of iterations and computational run-time. 

Careful selection of the cost function was crucial to the success of this work, and interacted with the choice of the aforementioned algorithms, particularly the local optimizer. Throughout the paper, we define the fidelity of the gate in terms of the Hilbert-Schmidt inner-product between the target unitary gate operation $U_0$ and the unitary operation $U$ generated by the Hamiltonian at a given step of the optimization (restricted to the logical subspace). Specifically, the fidelity is defined to be
\[
F(U_0, U) = |{\hs{U_0}{U}}|
\]
with
\[
\hs{U_0}{U} = \frac{\Tr (U_0 ^\dagger U)}{N},
\]
where $N$ is the dimension of the logical space (4 for two-qubit gates). This fidelity can be interpreted as a sort of lower-bound average fidelity of the gate.

To be specific about the iterative numerical process, at each step we generated $U$ from a vector corresponding to lattice parameters and calculated $F(U_0, U)$. Numerically, we found that minimizing the function $1 - F^2$, rather than $1-F$, gave superior performance. In the case of the algorithm given in Ref. \cite{spowell_bobyqa_2009}, the reason for this is clear: the algorithm assumes a quadratic cost function. However, we found that even with algorithms designed for linear cost functions (e.g. \cite{spowell_direct_1998}), convergence was much slower than for the quadratic cost function.

Finally, in order to ensure that $U$ has the same global phase as $U_0$ (this is for aesthetic purposes, as $F(U_0, U)$ is invariant under multiplication by a global phase), we placed a cost on the phase of the matrix element $u_{1,1}$. We found this to be most efficiently implemented by adding the term $\sin(\arg(u_{1,1}))^2$ to the cost function. This function is quadratic when perturbed about zero, is non-negative, and is symmetric about $n \pi$ for all $n \in \mathbb{Z}$, making it an ideal candidate function. We verified that the introduction of this additional cost both yielded a $U$ with appropriate phase (see Fig. 2 in the main text) and did not result in a decreased fidelity compared to optimization without this constraint.

\end{document}